\documentclass[10pt]{iopart}
\usepackage{graphicx}
\usepackage{bm} 
\begin{document}

%--------------------------------------------------------------
\title{Correlations between political party size and voter memory: A statistical analysis of opinion polls}
%--------------------------------------------------------------

\author{Christian A. Andresen}
\ead{christian.andresen@ntnu.no}
\address{Department of Physics, Norwegian University of Science and
Technology, N--7491 Trondheim, Norway}

\author{Henning F. Hansen}
\ead{henning.hansen@ntnu.no}
\address{Department of Physics, Norwegian University of Science and
Technology, N--7491 Trondheim, Norway}

\author{Alex Hansen}
\ead{alex.hansen@ntnu.no}
\address{Department of Physics, Norwegian University of Science and
Technology, N--7491 Trondheim, Norway}

\author{Giovani L. Vasconcelos}
\ead{giovani@lftc.ufpe.br}
\address{Laborat\'{o}rio de F\'{i}sica Te\'{o}rica e Computacional, Departamento de F\'{i}sica, Universidade Federal de Pernambuco, 50670-901 Recife, Brazil}

\author{Jos\'{e} S. Andrade Jr}
\ead{soares@fisica.ufc.br}
\address{Departamento de F\'{i}sica, Universidade Federal do Cear\'{a}, 60451-970 Fortaleza, Cear\'{a}, Brazil}

%--------------------------------------------------------------
\date{\today}
%--------------------------------------------------------------
\begin{abstract}
This paper describes the application of statistical methods to
political polling data in order to look for correlations and memory
effects. We propose measures for quantifying the political memory
using the correlation function and scaling analysis. These methods
reveal time correlations and self-affine scaling properties
respectively, and they have been applied to polling data from
Norway. Power-law dependencies have been found between correlation
measures and party size, and different scaling behaviour has been
found for large and small parties.
\end{abstract}

%89.65.-s   Social and economic systems.
%05.45.Tp   Time series analysis
%01.75.+m   Science and society

%--------------------------------------------------------------
\maketitle
%--------------------------------------------------------------

\section{Introduction}

\begin{quote}
\textit{``It has been said that democracy is the worst form of government except all the others that have been tried.''}

\quad\quad\quad\quad\quad\quad\quad\quad - Sir Winston Churchill
\end{quote}

The concept of democracy dates back to the Ancient Greeks and their
political and philosophical ideas. In a true democracy the power of
the government is vested in the people, in one way of another, via
the realization of free elections at regular time intervals. Owing
to its including nature, democracy tends to act as a stabilizing
force within a society, usually preventing the implementation of
catastrophic ideas and policies and ensuring peace and prosperity.
Studies show that democratic nations more seldom go to war
\cite{Weart}, have fewer civil wars \cite{Hegre}, have higher
economic growth, manage their resources in a better way
\cite{Carnegie1}, and provide better education and health care systems
than less democratic nations do.

A democracy can be seen as a system of interacting people exchanging
political opinions and ideas. In such a highly complex system one
should expect  nontrivial phenomena to emerge spontaneously. For
instance, scaling \cite{soares1,soares2} and universality \cite{fotrunato} have
recently been observed in the distribution of votes received by
candidates in parliamentary elections. By the same token, the time
evolution of political preferences within a given society
is expected to display scaling behaviour
that may indicate, for instance, the existence of correlations and
memory effects. Opinion variations have also been studied within the 
framework of modern network theory \cite{travieso,moreira}, and a complex behaviour have been found.
However, the fact that election
time series are usually very short,  makes it difficult to perform
meaningful scaling analysis. On the other hand, many modern democracies
have detailed and frequent political polling data over several decades
which result in longer time series where more reliable analysis can be
carried out. These polling data are  well suited for finding the
political memory of the voter base for different political
parties.

In this paper, we perform an empirical analysis of Norwegian polling
data over the last 30 years and find a power-law relationship between
the correlation time and party size. We also compute the Hurst
exponent of the time series of party  preferences and find
different exponents for large and small parties.  In particular, we
observe that large (small) parties have a Hurst exponent smaller
(greater) than $1/2$, thus indicating anti-persistent
(persistent) trend. To the best of our knowledge, this is the first time
that such scaling laws have been found in electoral time series. (A
trend analysis of German elections has recently been performed by
Schneider and Hirtreiter
\cite{schneider} but in a different context.)

The paper is organized as follows. In Sec.~2 we briefly discussed the
statistical methods we shall use for detecting correlation and memory
effects. In Sec.~3 we describe the Norwegian polling data on which we
shall apply our correlation and scaling analyses, with the results
being presented in Sec.~4. Finally, in Sec.~5 we summarize our main
conclusions.

%--------------------------------------------------------------
\section{Methodology}

We present two different methods for quantifying the political memory, based on the correlation function and scaling analysis. The scaling analysis is performed by using the Average Wavelet Coefficient (AWC) method. Both methods are described in detail below.

\subsection{Correlation function}

The auto-correlation function $C_{XX}(t_{\Delta})$ \cite{cross_corr} for a discrete time series $X_t$ is defined by

\begin{equation}
C_{XX}(t_{\Delta})  =  \frac{1}{(N-t_{\Delta})\sigma_{X}^{2}} \sum_{t=1}^{N-t_{\Delta}} [ (X_t - \mu_X )(X_{t+t_{\Delta}} - \mu_X )],
\label{eq:xcorr}
\end{equation}
where $N$ is the number of points in the time series, $\mu_X$ is the average value and $\sigma_X$ is the standard deviation for $X_t$.

The auto-correlation function describes how the time series, $X_t$, 
correlates with itself over different time scales, hence describing the 
memory of the system. The time when the function first turns negative 
is denoted here as $t_0$, and is the time at which the series first becomes 
uncorrelated. A large $t_0$ indicates a long memory, while a small $t_0$ 
indicates a short memory. Note that in noisy time series the auto-correlation 
function may briefly turn negative before turning positive again over a long 
interval. In such cases this first negative 'dip' may have to be ignored as 
the data shows an underlying correlated trend. 

The auto-correlation function can also be used to define the integrated correlation time, $\tau$, given by
\begin{equation} 
\tau^2 = \sum_{t_{\Delta} =1}^{\infty} t_{\Delta}^2 C_{XX}(t_{\Delta}),
\label{eq:integrated_time}
\end{equation} 
where $C_{XX}(t_{\Delta})$ is defined by Eq. \ref{eq:xcorr}. $\tau$ is a measure of the persistence of the correlation, a large $\tau$ means that the signal have been highly correlated for a long time, and thus have a long memory.

\subsection{Wavelet analysis and Hurst exponent}

To look for scaling properties in the polling data for each political
party, we focus on the conditional probability density $p(t,y)$,
where $y$ is the electoral preference (in percent). This is the
probability that a given political party has a preference passing within $dy$ of the preference
$y$ when it passed through $y=0$ at time $t=0$. This probability
density shows the invariance

\begin{equation}
\lambda^{H} p(\lambda t, \lambda^{H} y)=p(t,y),
\label{eq:probability_density}
\end{equation} 
where $H$ is the Hurst exponent \cite{hurst}.  A Hurst exponent within the range $1/2 < H < 1$ implies a persistent,
trend-reinforcing series, while for $0 \leq H < 1/2$ one has an
anti-persistent time series. A pure random walk gives $H=1/2$.

To estimate the Hurst exponent, we perform a wavelet analysis
\cite{simonsen} of the polling data for each political party. A
wavelet analysis is the preferred way to estimate the Hurst exponent
when having small data sets \cite{simonsen}. When the average
wavelet-coefficients are plotted as a function of time scale with
log-log axis, the best fitted slope is given by $H+1/2$.

\section{The Data}

In order to illustrate the ideas for political memory described above,
we shall apply them to monthly polling data taken in Norway over the
last 30 years.  Before presenting the data, it is perhaps instructive
to describe briefly the Norwegian political landscape.

Norway is a constitutional monarchy \cite{smk1} with a parliamentary
democratic multi-party system. The Norwegian King has symbolic power,
and his functions in political sense are ceremonial.  After an election 
the monarch will ask the leader of the parliamentary block
that has the majority in the elected national assembly ``Stortinget''
to form a council. ``Stortinget'' currently has 169 members
\cite{smk2}. These members are elected for 4-year terms from 19
administrative regions. The seats in ``Stortinget'' are distributed
based on a system of proportional representation.

\begin{figure}[h!]
\begin{center}
$\begin{array}{c}
\resizebox{0.6\columnwidth}{!}{\includegraphics[angle=-90]{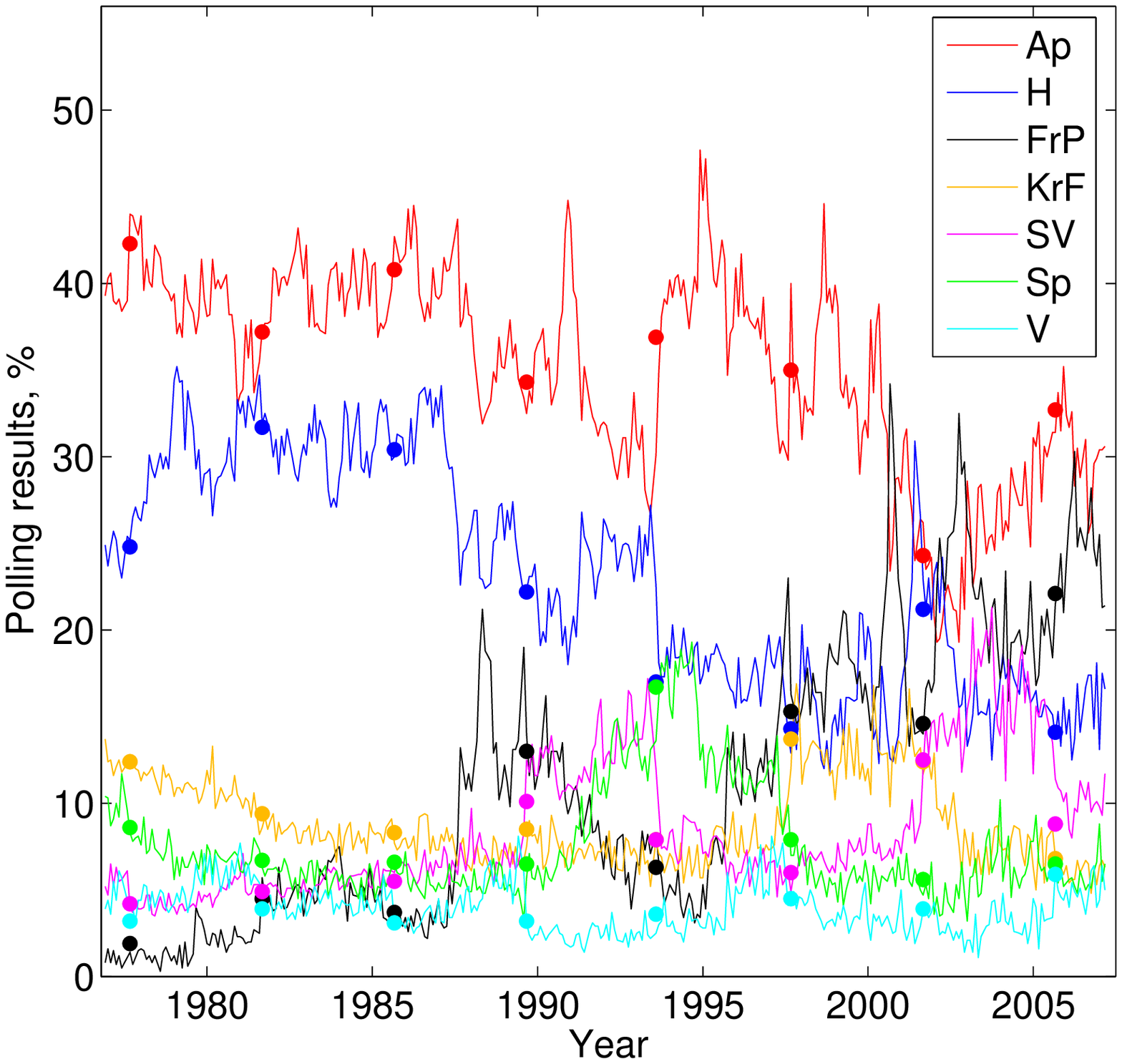}}\\
\resizebox{0.4\columnwidth}{!}{\includegraphics[angle=0]{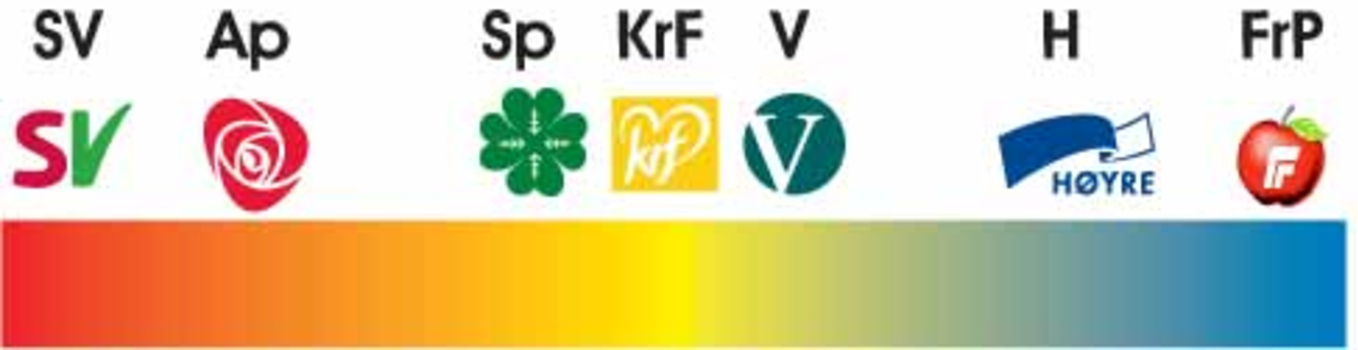}}
\end{array}$
\end{center}
\caption{Upper panel: Monthly polling results for the period from December 1976 to March 2007. The solid dots represents parliamentary election results. Lower panel: Simplified one-dimensional representation of the Norwegian political landscape along the traditional left-right axis.}
\label{fig:norway}
\end{figure}

The Norwegian political landscape for the last 30 years has been
dominated by 7 major political parties, on average jointly receiving
98 $\%$ of the total votes. These can be ordered on a simplified
one-dimensional axis with two parties on the left, Socialist Left
Party (SV) and Norwegian Labour Party (Ap), two parties on the right,
Progressive Party (FrP) and Conservative Party (H) and 3 parties in
the centre, Centre Party (Sp), Christian Democratic Party (KrF) and
Liberal Party (V). This is illustrated in the lower panel of Figure \ref{fig:norway}.

In our analysis of political memory effects we shall use monthly
Norwegian polling data from ``Synnovate Norway'' taken from December
1976 to March 2007. The data is shown in the upper panel of
Figure
\ref{fig:norway}.  The number of people interviewed in
each survey is of the order of 1000, and the polling data
therefore have a considerable degree of uncertainty.  The ranges of
the 95 $\%$ confidence intervals for the polling average of the
various parties are listed in Table
\ref{tab:parties}. In spite of their uncertainty, the 
polls performed remarkably well in predicting the actual outcome
of the elections. This can be seen in Figure
\ref{fig:norway} where the the results from the parliamentary elections
held every four years are also plotted as solid circles. This 
good agreement further reinforces the fact that polling data are indeed suitable
for the analysis of the true electoral dynamics.

\section{Results}

Here we shall apply the auto-correlation and the AWC analyses in
order to probe correlations and memory effects in the Norwegian
polling data shown above. 

\subsection{Auto-correlation analysis}

\begin{figure}[ht!]
\begin{center}
$\begin{array}{cc}
\resizebox{0.4\columnwidth}{!}{\includegraphics[angle=0]{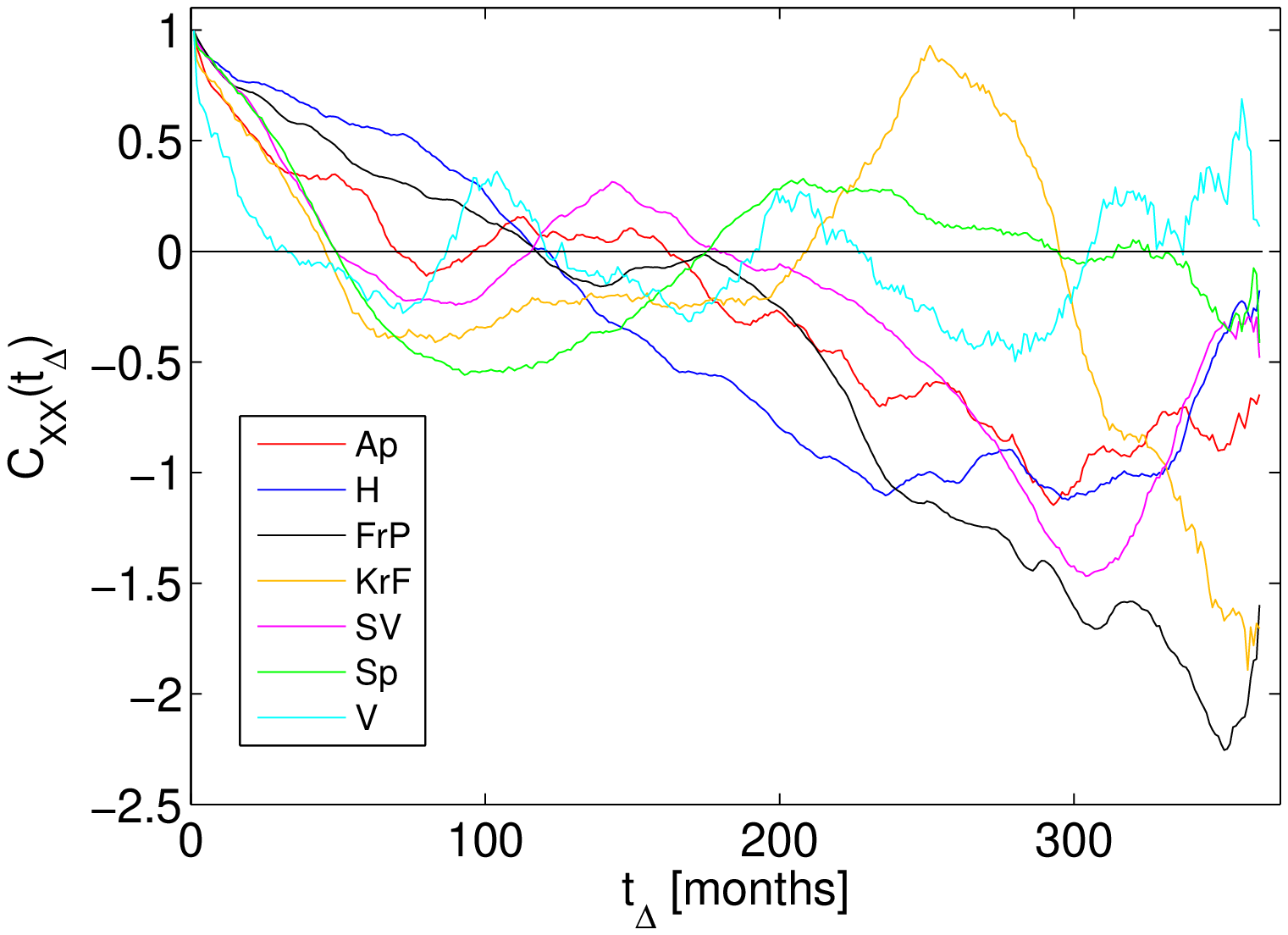}}&
\resizebox{0.4\columnwidth}{!}{\includegraphics[angle=0]{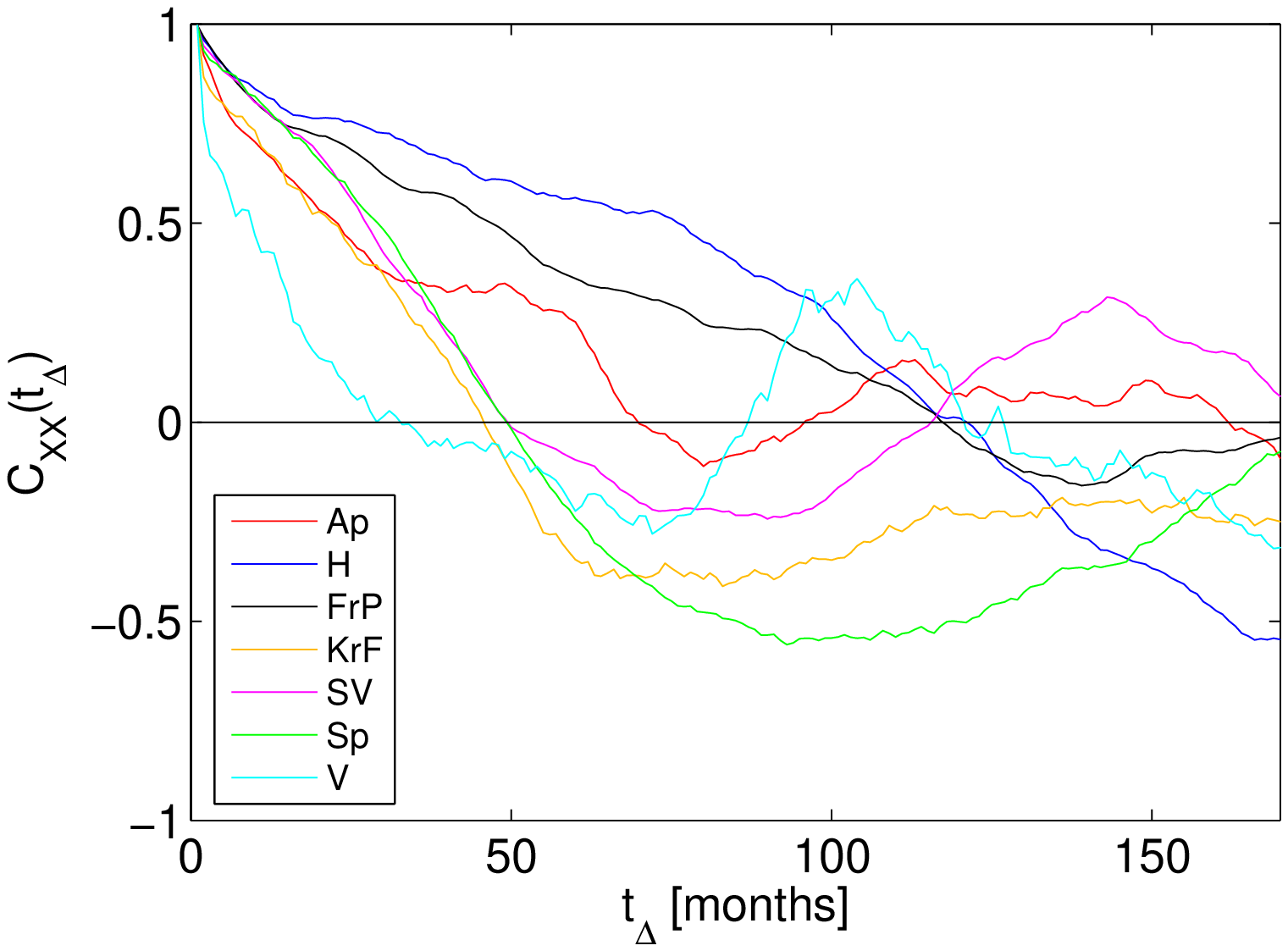}}
\end{array}$
\end{center}
\caption{Left: The auto-correlation function, $C_{XX}(t_{\Delta})$, given in Eq. \ref{eq:xcorr}, as function of time delay, $t_{\Delta}$, for the 7 major political parties in Norway. Right: The auto-correlation function with focus on the small-scale behaviour.}
\label{fig:corr}
\end{figure}

In our analysis, $X_t$ is the polling results for a given political
party at time $t$ and we  use Eq. (\ref{eq:xcorr}) in
order to look at the auto-correlation function. This function is
usually applied to stationary ensembles, however this is not the case
for political polling data on large time scales as seen in the left
panel of figure \ref{fig:corr}. We therefore focus on the correlation
function for short time scales, shown in Figure \ref{fig:corr}, and
turn our attention to the integrated correlation time given in
Eq. \ref{eq:integrated_time}.

\begin{table}[h]
\begin{center}
\begin{tabular}{l rccrrrr}
\hline
  & $\overline{X}_t$ & $\sigma(X_t)$ & err($\overline{X}_t$) & $\tau$   & $\sigma(\tau)$   & $t_{0}$ & $\sigma(t_{0})$ \\ 
\hline
Ap                         & 35.2  & 5.8 & 3.0  & 154 &  5.7 &   67 &  2.70  \\
$\textrm{Ap}_{\textrm{sec}}$                   &       &     &      & 320 &      &  151 &        \\  
H                          & 22.9  & 6.4 & 2.6  & 385 &  4.0 &  113 &  1.12  \\
FrP                        & 11.0  & 7.9 & 2.0  & 310 &  4.9 &  109 &  0.60  \\
KrF                        &  8.9  & 2.4 & 1.9  &  81 &  2.8 &   42 &  0.90  \\
SV                         &  8.5  & 3.9 & 1.8  &  97 &  2.4 &   46 &  1.16  \\
Sp                         &  7.6  & 3.3 & 1.8  &  98 &  2.4 &   46 &  0.96  \\
V                          &  3.9  & 1.3 & 1.4  &  34 &  4.5 &   31 &  4.20  \\ 
\hline
\end{tabular}
\caption{Table showing the polling average, $\overline{X}_t$, the standard deviation of the polling data, $\sigma(X_t)$, the $\pm$ range of the 95 $\%$ confidence interval of the polling data, err($\overline{X}_t$), the integrated correlation time, $\tau$, the standard deviation of the integrated correlation time, $\sigma(\tau)$, the first zero crossing of the correlation function, $t_0$ and the standard deviation of the first crossing of the correlation function $\sigma(t_0)$. The values for $\textrm{Ap}_{\textrm{sec}}$ is for the second zero-crossing value for Ap.}
\label{tab:parties}
\end{center}
\end{table}

To give meaning to $\tau$ in our case, and thereby making is useful 
as a measure of political memory, we cut the sum at the first 
zero-crossing point $t_0$ when the correlation
function for the first time shows an anti-correlated behaviour. Since
the data is related to a given uncertainty, the first crossing time
and the integrated correlation time are also associated with an
uncertainty. We have calculated the standard deviation of $t_0$ and
$\tau$ for all the parties by looking at an ensemble of 1000 samples
where uniform noise has been added to the data. The range of the noise
for each party was set to the range of the 95 $\%$ confidence
interval, as indicated in Table \ref{tab:parties}.

\begin{figure}[ht!]
\begin{center} 
\resizebox{\columnwidth}{!}{\includegraphics[angle=0]{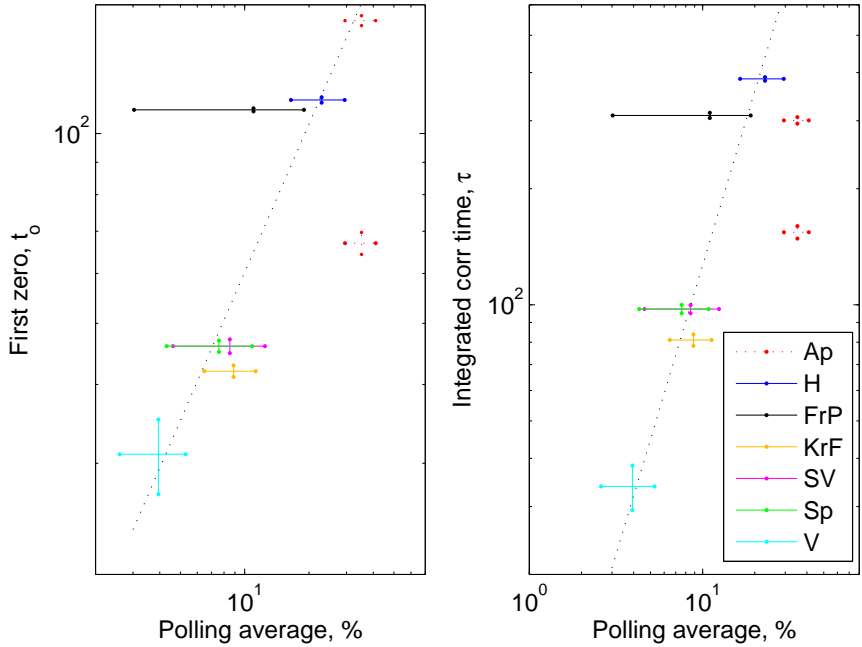}}
\caption{First anti-correlation time, $t_0$, and integrated correlation time, $\tau$, vs. polling average, $\overline{X}_t$, for the investigated parties. The data is plotted in a log-log plot and fitted with a best fit power-law functions $\tau \propto \overline{X}_t^{\beta_{t_o}}$ and $t_0 \propto \overline{X}_t^{\beta_{\tau}}$. For the first zero-crossing time $\beta_{t_0} = 0.78$ and for the integrated correlation time $\beta_{\tau} = 1.50$ }
\label{fig:loglog_corr}
\end{center} 
\end{figure}

In Figure \ref{fig:loglog_corr} we plot $t_0$ and $\tau$ as a function
of the polling average for each party. These plots indicate that both
the first crossing time and the integrated correlation time increase
with the size of the party (polling average). In both plots a
power-law dependence have been found to fit the trend of the data. The
data points for Ap lies outside this trend. However if we look at the
auto-correlation function in Figure \ref{fig:corr} we see that Ap is
the only party that shows a brief anti-correlated behaviour before it
again becomes correlated. The next interval of anti-correlated
behaviour is persistent and in accordance with the other parties. If
we use the second zero-crossing of Ap the data point for this party
follows the fitted trend for $t_0$. We have found the best fit to be
$\tau \propto \overline{X}_t^{1.50}$ and $t_0 \propto
\overline{X}_t^{0.78}$. We interpret both $\tau$ and $t_{0}$ as a
measure of the political memory for each party's
voter base. Our results thus show that the larger the size of a party 
the longer the  political memory of its voter base.
%-------------------------------------------------------------- 
%-------------------------------------------------------------- 

\subsection{AWC analysis}

\begin{figure}[ht!]
\begin{center}
$\begin{array}{cc}
\resizebox{0.45\columnwidth}{!}{\includegraphics[angle=0]{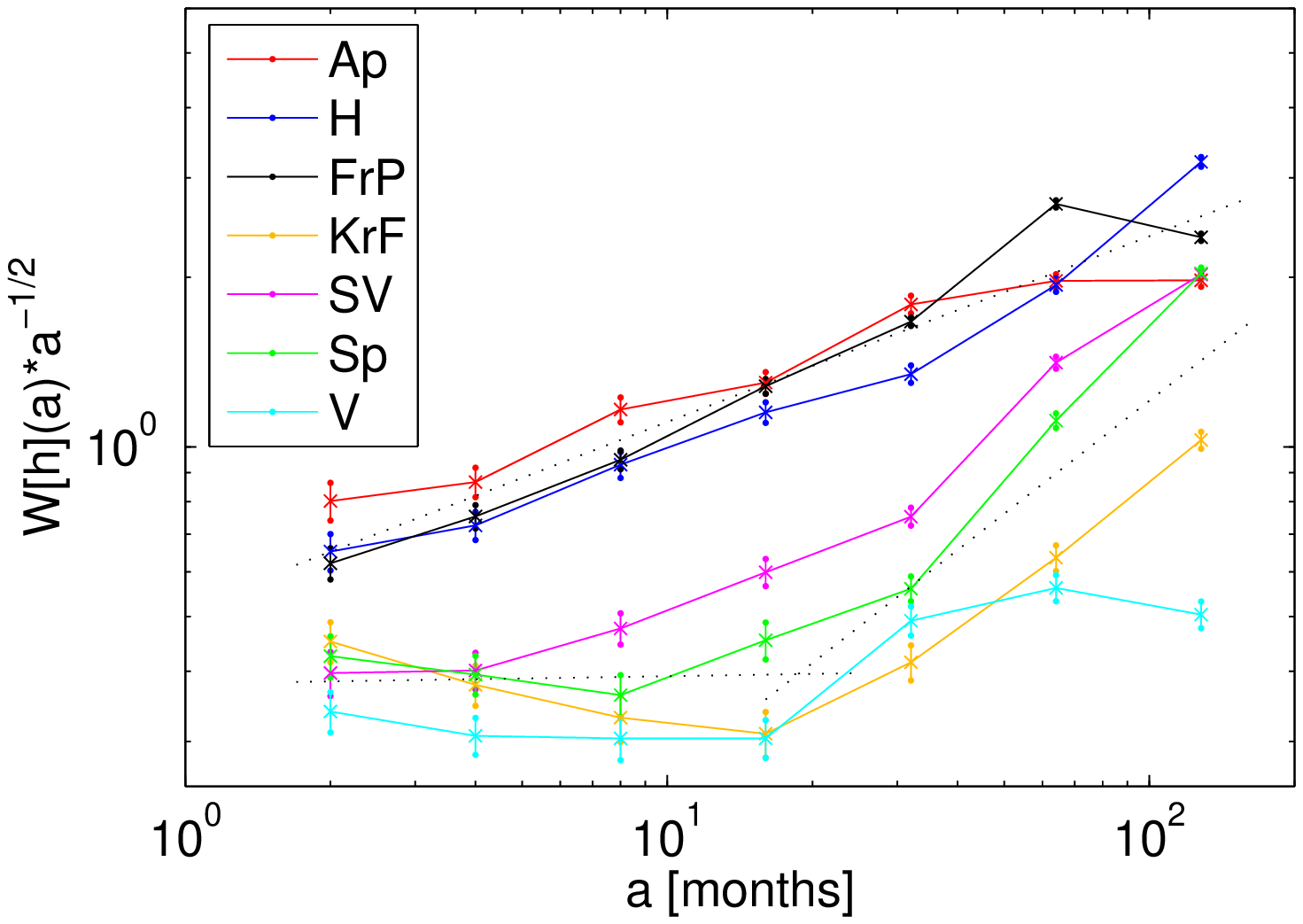}}&
\resizebox{0.45\columnwidth}{!}{\includegraphics[angle=0]{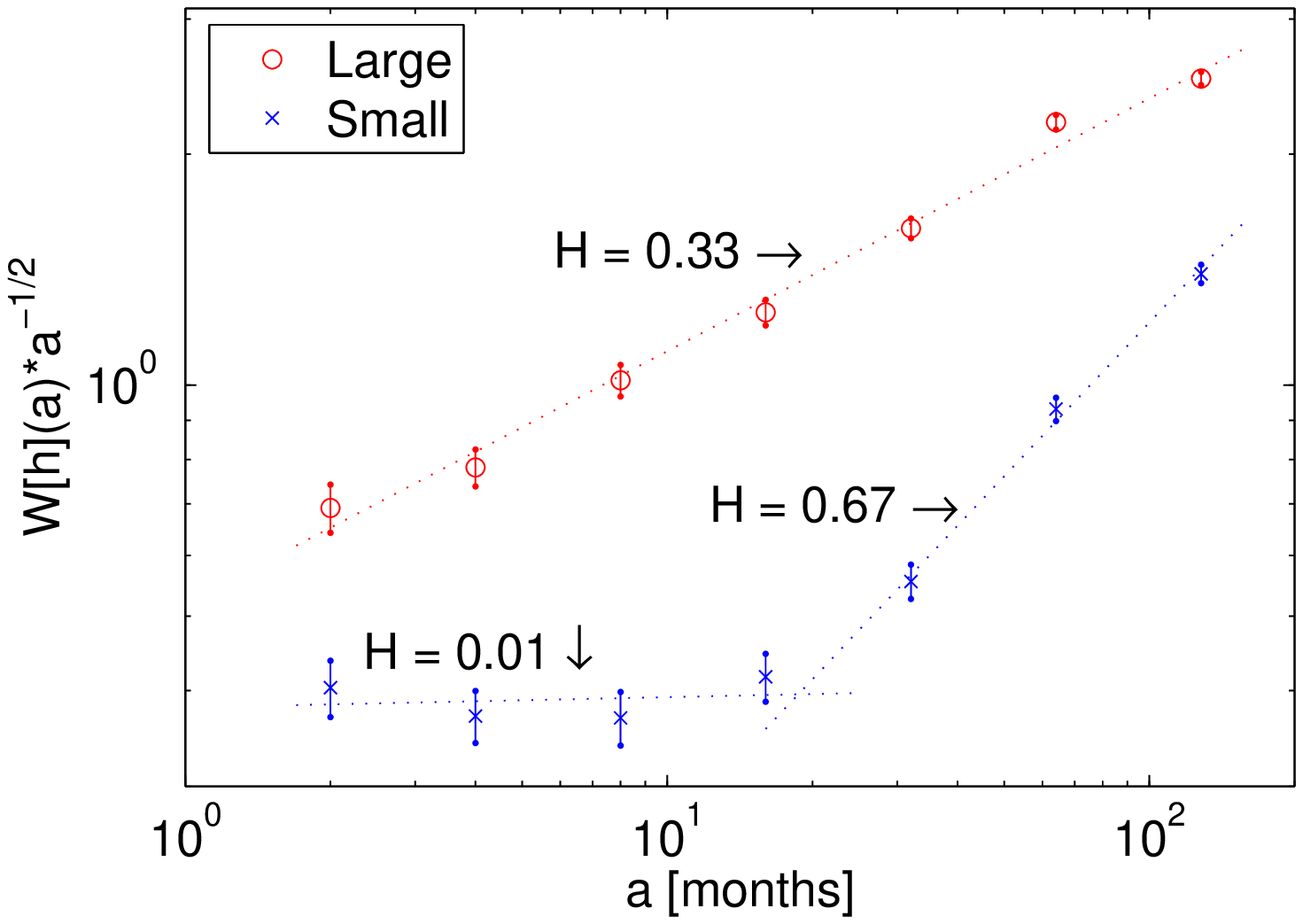}}
\end{array}$
\end{center}
\caption{Left: The average wavelet-coefficients on a given time scale, $a$, scaled with $a^{-1/2}$ as a function of the time scale. Right: The average wavelet-coefficients grouped into large and small political parties. The results are averages for all the parties in each group.}
\label{fig:awc}
\end{figure}

In the left panel of figure \ref{fig:awc} we have plotted the AWC results, 
scaled by $a^{-1/2}$ where $a$ is the time scale, as a function of the time 
scale for each political party. The error bars plotted in Figure \ref{fig:awc} 
indicate $\pm$ one standard deviation for the wavelet coefficient based on 
1000 samples where each sample have been added flat noise in the same fashion 
as in the calculation of the error bars for $\tau$ and $t_0$ above.

Looking at the polling average for all the parties (left panel of
Figure \ref{fig:awc}), we can
separate them into two groups. One group of large parties (Ap, H and
FrP), and one group of small parties (SV, Sp, KrF and V). From Table
\ref{tab:parties} one can see that all  parties within the same group
have similar polling averages $\overline{X}_t$. They also have first
crossing times $t_0$ in the same range. As a result we have averaged
the AWC results for all the parties within each of the two
groups, and the results can be seen in the right panel of Figure 
\ref{fig:awc}. The best
linear fit on log-log scale is also inserted in the left panel of 
Figure \ref{fig:awc}. The averaged results suggest that the large
parties have a Hurst exponent of 0.33 over the entire range that can
be analysed by our sparse data, thus indicating anti-persistent
behaviour in voter preferences for these parties. Since the larger
parties are more likely to win the elections and form a Government,
the anti-correlated behaviour seems to support the view that it is hard
to remain in power for successive terms. Furthermore, the other
parties will most certainly target voters from the leading party's
base hoping to win their preference, and this competition may
contribute for the anti-persistence seen in the polling data for large
parties. As we will see below, smaller parties are less susceptible
to this effect.

As seen in both panels of Figure \ref{fig:awc}, the averaged wavelet
coefficients for smaller parties show a shift in behaviour at about 1
to 3 years, or about 10 to 40 months. They cross over from a Hurst
exponent close to zero for small time scales, indicating a highly 
anti-persistent trend, to a Hurst exponent of 0.67 for larger time
scales indicating a correlated behaviour. The scale at which the
crossover occurs is in the same range as the correlation measure $t_0$
for the smaller parties. For the larger parties $t_0$ is so large that
we can not see any crossover in the wavelet analysis. This may
indicate that the ``recollection time'' or memory  in the smaller parties
is of the order of 2 years, whereas for the larger parties it is
considerably longer and possibly on the scale of decades.

\section{Summary and Conclusions}

In this paper we have put forward two different methods for measuring
political memory, namely, correlation analysis and scaling
analysis. To illustrate these measures, we have carried out an
empirical analysis of monthly polling data for the major political
parties of Norway over the last 30 years. It has been found that the
integrated correlation time $\tau$ and the time of first occurrence of
anti-correlation $t_0$ grow as a function of party size
$\overline{X}_t$, and both have been found to display a
power-law dependence: $\tau \propto
\overline{X}_t^{1.50}$ and $t_0 \propto \overline{X}_t^{0.78}$.

The polling records for all the parties show self-affine properties,
and the Hurst exponent $H$ have been calculated using the AWC
method. It has been shown that the polling records show
anti-correlated behaviour with $H = 0.33$ for large parties. For
smaller parties a crossover from a highly anti-persistent trend
$H=0.01$ to a correlated behaviour with $H=0.67$ has been found.

Taken in conjunction, the results from the correlation and
scaling analyses show that the voter bases for the large parties have
inherently long memories and this seems to lead to an mean-reverting trend
in their electoral preferences. On the other hand, the smaller
parties have little short-term memory and a highly anti-persistent
behaviour at the small time scales, however for longer times they show a 
crossover to a persistent trend  when the data sets become
anti-correlated. Unfortunately, the time series are too short to show
any crossover for larger parties.

The discussion above indicates that smaller parties, which are
often perceived to fight for special interests, have over long time
scales more persistent voters than the larger parties do. They
have more rapidly varying polling data on small time scales, but on
the longer time scales their voters are more loyal. On the other
hand the large parties, often perceived as mass parties that appeal to
a large fraction of the population, have slower varying polling data
on short time scales in comparison to small parties, but on longer
time scales their voters are less  loyal.

\end{document}